\documentclass[superscriptaddress,twocolumn,nofootinbib,amsmath]{revtex4}

\usepackage{graphicx}% Include figure files
\usepackage{dcolumn}% Align table columns on decimal point
\usepackage{bm}% bold math

\newcommand{\be}{\begin{equation}}
\newcommand{\ee}{\end{equation}}
\newcommand{\bea}{\begin{eqnarray}}
\newcommand{\eea}{\end{eqnarray}}
\newcommand{\pat}{\partial}

\newcommand{\half}{\frac{1}{2}}

\newcommand{\ket}{\rangle}

\newcommand{\comment}[1]{}

\begin{document}

\title{Brane Decay from the Origin of Time}

\author{Shinsuke Kawai}
%\email{shinsuke.kawai@helsinki.fi}
\affiliation{Helsinki Institute of Physics, P.O. Box 64, FIN-00014
University of Helsinki, Finland} 
\affiliation{ Yukawa Institute for
Theoretical Physics, Kyoto University, Kyoto 606-8502, Japan}
\author{Esko Keski-Vakkuri}
%\email{esko.keski-vakkuri@helsinki.fi}
\affiliation{Helsinki Institute of Physics, P.O. Box 64, FIN-00014
University of Helsinki, Finland} 
\affiliation{Department of Physical Sciences, P.O. Box 64, FIN-00014 
University of Helsinki, Finland}

\author{Robert G. Leigh}
%\email{rgleigh@uiuc.edu}

\author{Sean Nowling}
%\email{nowling@students.uiuc.edu}
\affiliation{Department of Physics,
University of Illinois
1110 W. Green Street, Urbana, IL 61801, U.S.A.}

\date{\today}

\begin{abstract}
We present a novel scenario where matter in a spacetime originates from a
decaying brane at the origin of time. The decay could be considered
as a ``Big Bang''-like event at $X^0=0$.
The closed string interpretation is a
time-dependent spacetime with a semi-infinite time direction, with
the initial energy of the brane converted into energy
flux from the origin. The open string interpretation can be viewed as
a string theoretic non-singular initial condition.
\end{abstract}

\pacs{}
\keywords{}
\maketitle

%\section{Introduction}
{\em Introduction -- } Perhaps the most ambitious problem in
cosmology is the question of the initial conditions of the Universe.
In this paper we present a completely new scenario: brane nucleation
and decay from the origin of the spacetime. Various proposals for
initial conditions have been formulated in different
frameworks (see \cite{Carroll:2004pn} for a recent discussion). The
most famous is the no-boundary proposal, where an expanding universe
emerges from an instanton in the Euclidean geometry, thus removing
the boundary from the spacetime manifold \cite{nobdry}.
Alternatively, it has been speculated that a preceding long pre-Big
Bang phase \cite{Gasperini:2002bn} ended in a Big Crunch, with the
Universe re-emerging into a Big Bang. It has been hoped that this
so-called Big Bounce could be reliably described, once the true
quantum gravitational effects would be properly understood, perhaps
in the context of string theory.  This proposal then motivated a
variant \cite{Khoury:2001wf} where the Big Bounce was attributed to
a collision of branes moving in a higher dimensional space, either
as a singular event or in repeated cycles. The associated spacetimes
were modeled in string theory as Lorentzian boost orbifolds,
but as of yet it is still an open question whether a detailed
understanding of the associated stringy effects \cite{Durin:2005ix}
could lend support to the bounce idea.

A related development is the AdS/CFT duality,
or more generally the open-closed duality of string theory.
Based on models of stacks of stable D-branes, one can
show a duality between the open string theory on the brane and
closed string theory in a higher dimensional spacetime. In the
bulk spacetime, the additional spacelike direction can be related
to renormalization group flow in the lower dimensional effective
field theory of open strings on the brane. The branes in question
have a timelike worldvolume. It was hoped that one could develop
analogous models by replacing the timelike branes with spacelike
branes \cite{Gutperle:2002ai};
the open string physics on the brane would give
rise to a time direction, so that the dual closed string
interpretation would be string theory in a higher
dimensional time-dependent spacetime.

Natural candidates for the
spacelike branes are unstable D-branes or D-brane -anti D-brane
configurations. The decay of these brane
configurations can be modeled in open string theory in different
ways, {\em e.g.} as
conformal field theory of the open string worldsheet. In this
case, the brane decay can be modeled by a rolling tachyonic background
\cite{Sen:2002nu} introduced into the action as an
exactly marginal
deformation. On the other hand, for the dual spacetime
interpretation one needs to find time-dependent (S-brane) solutions of
supergravity which are sourced by the associated unstable D-brane
configuration. This programme is hampered by several technical
difficulties on both sides. %On one hand, for example,
It is difficult to
construct generic exactly marginal tachyonic
deformations. %On the other hand,
Few examples of S-brane
spacetimes are known. Furthermore, string dynamics in such
backgrounds is technically very involved,
obscuring the
problem of identifying interesting scaling limits for simplified
dualities.

The prototype
scenarios for brane decay come in two basic variants. In the first
\cite{Sen:2002nu}, the brane decay has a finite time origin, but the
decay phase is preceded by a build-up phase, where the unstable
brane is first created by a carefully fine-tuned closed string
initial state. In fact, the whole creation-and-decay process is
completely time reflection symmetric. In the second variant
\cite{Larsen:2002wc}, the decay starts at past infinity and
continues through all time until future infinity.
The first variant has the virtue that the timescale of the decay is a
tunable parameter and the decay has a finite origin. But this is
achieved at the expense of
a highly fine-tuned creation phase. Alternatively, a prescription
for brane nucleation in an empty spacetime
has been proposed in \cite{Lambert:2003zr}.
The second variant has
the virtue of only having decay, but the downside is not having a
finite initial time and no tunable timescale.

We now propose a new model where the brane decay starts from a finite
time origin, has a tunable timescale for the decay, and no creation
phase. This opens up exciting possibilities for scenarios where the
unstable brane on one hand provides initial conditions which can be
phrased in terms of open strings on the brane and on the other hand
has a dual interpretation of a time-dependent spacetime with a
finite origin of time. The details have some flavor of the
no-boundary proposal in terms of involving physics in the complex
time plane, some flavor of the pre-Big Bang idea, and some flavor of
Lorentzian string orbifold models, the AdS/CFT duality and perhaps
even holography.
This paper presents the main features of our model, a more detailed
discussion will follow in \cite{KKLN}.

As this paper was prepared, other new ideas about
cosmological singularities were reported
in \cite{othercoolideas}.

%\section{The Model}
%\label{sec:model}
{\em The model -- }
We begin by recalling the basic pre-Big Bang idea: to resolve the initial
singularity to extend the
past history of the Universe, as sketched in Figure 1(a).

%%%%%%%%%%%%%%%%%%%%%%%%%%%%%%%%%%%%%%%%%%
\begin{figure}
\includegraphics[width=7cm, height=5cm]{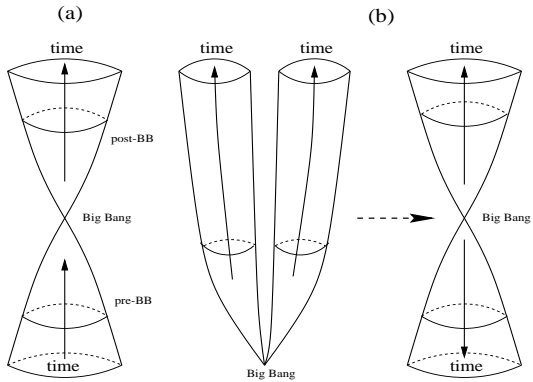}% Here is how to import EPS art
\caption{\label{fig:BBB_fig1}
The pre-Big Bang scenario (a), and the creation of two universes from the Big Bang,
which is interpreted as a spacetime ${Z}_2$ orbifold (b).
}
\end{figure}
%%%%%%%%%%%%%%%%%%%%%%%%%%%%%%%%%%%%%%%%%%

The concept of ``time" breaks down at the Big Bang
singularity. In Big Crunch / Big Bang scenarios one hopes to
continue time's arrow across, by a resolution mechanism \cite{Seiberg:2002hr}.
However, one can ask if time's arrow could be taken to
point in multiple directions from
the Big Bang. We could imagine several universes being created,
such as is suggested by Figure 1(b).
It might be that this allows for a different kind of resolution of
the initial singularity.
An additional ingredient which
can be added to such considerations is CPT. One
could postulate that the two Universes are simply CPT reflections
of one another. Then it would be possible to identify them under
the CPT reflection. An analogous proposal has been made before in
the elliptic interpretation of the de Sitter space \cite{EdS}.

Consider then the two main models for brane decay; for simplicity,
we focus on D-branes in bosonic string theory. The first variant, mentioned above,
is the full S-brane, associated in the boundary CFT of the open
string worldsheet with the spatially homogenous tachyonic deformation
\begin{equation}
T(X^0) = \lambda \cosh (X^0 /\sqrt{\alpha'})
\end{equation}
where the parameter $0\leq \lambda \leq \half$ controls the
lifetime of the brane: the brane exists for a finite time of the order
\begin{equation}
\Delta X^0 \approx 2 \ln (\sin (\pi \lambda )) \ .
\end{equation}
The decay starts at $X^0=0$ and
is preceded by a formation phase, where the brane forms out
of a carefully fine-tuned initial closed string configuration,
which can be mapped to the final state by time reversal. The
situation is
analogous to building a bomb by reversing its
explosion. From the target space point of view, this appears
unnatural. It is possible to isolate the decay phase by
considering instead a deformation $T(X^0)=\lambda e^{X^0}$, corresponding to the second variant, the
half-S-brane. Now
the parameter $\lambda$ has no physical meaning as it can be absorbed into
a shift of the origin of the time coordinate.

The full
S-brane in Lorentzian spacetime is related by a Wick rotation
$X^0 \mapsto iX$ to
an infinite periodic array of smeared D-branes in Euclidean
space.
The tachyonic deformation becomes periodic
$\lambda \int_{\pat \Sigma} \cos (X/\sqrt{\alpha'})$
and alters the
boundary conditions at the endpoint of the open string.
Increasing $\lambda$ from initial value $0$ to a
final value $1/2$ smoothly deforms  the open string boundary conditions
from Neumann to Dirichlet, so that the open strings become
attached to a periodic array of D-branes
localized at $X=2\pi\sqrt{\alpha'}(n+\half)$, with
$n\in {Z}$. Conversely, as $\lambda$
is decreased from $1/2$, the
branes are smeared over the spacelike direction
transverse to the branes. Wick rotating back to the original
Lorentzian direction, the smearing corresponds to slowing down the
decay, with the lifetime $\Delta X^0$
related to the smearing of the branes.
Ref. \cite{Gaiotto:2003rm}
proposed that generic, suitably symmetric,
D-brane configurations in the
complexified time plane correspond to time-dependent closed string
backgrounds.
The D-brane configurations act as sources for the spacetime fields
and backreact to the geometry.
It is of great interest to find spacetime solutions which are
sourced by the brane arrays. Additional discussion and some
example solutions can be found eg. in \cite{Jones:2004rg}.

Let us now return to the
spacetimes of Figure 1 with the origin of time $X^0$ at the Big
Bang. Could one consider brane decay in such backgrounds?
In the case of (a), if the Crunch and the Bang are
time reverses of each other, one could attempt to construct
a full S-brane centered at the Big Bang. However, the spatial
slices pass through zero size, and one would need to take this
into account. For example, if the spacetime is modeled as a
Misner space,
one could attempt to construct a full brane solution where the
tachyon rolls with respect to the Misner time rather than with
respect to the Minkowski time.
In the case of (b), if the two branches of the spacetime emanating
from $X^0$ are mapped to each other under T-reversal,
it should be again be possible
to introduce the full S-brane.
But now there is an additional complication because
the spacetime is not globally time-orientable. Let us simplify the
problem by not requiring that the spatial sections undergo
expansion.
As a simple toy model, one can consider the
Lorentzian orbifold ${ R}^{1,d}/{ Z}_2 \times M_D$,
where ${ Z}_2$
acts as a (C)PT-reflection
\begin{equation}
(X^0,X^1,\ldots ,X^d)\mapsto (-X^0,-X^1,\ldots -X^d)
\end{equation}
in ${ R}^{1,d}$, and $M_D$ is a Euclidean space of
dimension $D=25-d$ (this would be replaced by $9-d$ in the superstring).
%In the above,
A priori $0\leq d\leq 25$. This
orbifold has been investigated in \cite{Balasubramanian:2002ry} and
\cite{Biswas:2003ku} in bosonic string
and type II superstring theories. The quantization
of untwisted and twisted sectors is
straightforward.
It was found that ghosts
are absent at tree level. In supersymmetric theory, the partition
function vanishes just as in the corresponding Euclidean orbifold.
There are no closed causal curves after the
proper definition of the time function on the fundamental domain.
The covering space
is then a Minkowski space, except that the time's arrow points in
opposite directions on the two half-spaces, as in Figure 1(b).
(If $d>0$, the spatial slice at $X^0=0$ is reduced in half due to
the P-reflection). Moreover, there is no dangerous backreaction
as the stress tensor vanishes everywhere after the Big Bang
initial slice.

Since the tachyonic deformation on the open string worldsheet is
symmetric with respect to the time reversal, the
full brane survives the orbifold ${ Z}_2$ identification.
Consider the closed string boundary state
description of the full brane. The starting point of the boundary
state construction is
the Wick rotation to the Euclidean
signature. The boundary state $|B\ket$ can be written in terms
of the underlying $SU(2)_L\times SU(2)_R$ symmetry with the
generators
%\be
$
  J^\pm = e^{\pm iX/\sqrt{\alpha'}} \ ; \ J^3 = i\partial X \ ,
$
%\ee
as a linear combination of Ishibashi states $|j,m,n\rangle
\rangle$ labeled by the $SU(2)$ quantum numbers.
The ${ Z}_2$ reflection $X\mapsto -X$ acts
on the $SU(2)$ generators
by
\begin{equation}
\label{Z2}
   J^\pm \mapsto J^\mp \ ; \ J^3 \mapsto -J^3  \ .
\end{equation}
The boundary state is symmetric with respect to interchanging $m,-m$,
so it is invariant under (\ref{Z2}).
The same will be true after the inverse Wick rotation back to
Lorentzian signature. But there is an important subtlety
involved, first discussed in \cite{Biswas:2003ku}. When the
boundary state $|B\ket$ is expressed in the standard oscillator and
momentum basis, it is a linear superposition of on-shell
closed string states $|\psi_i, k_i\ket$ with $k_i^2=-M_i^2$.
But, after the inverse Wick rotation, the ${ Z}_2$
reflection acts in time, so it also acts on center-of-mass energy
as $k^0\mapsto -k^0$. Therefore, the on-shell
states $|\psi_i , k_i\ket$ in the sum must be
doubled to reflect the symmetry of the choice of branch $k^0 =\pm
\omega_{\vec k}$:
\begin{equation}
  |\psi_i ,k_{iE}\ket \mapsto \left( \begin{array}{c}
  |\psi_i , k_i^0 ,\vec k_i \ket \\
  |\psi_i , -k_i^0 ,-\vec k_i \ket  \end{array} \right) \ .
\end{equation}
For more discussion, see  \cite{Biswas:2003ku}
and \cite{KKLN}.

The physical interpretation of the resulting boundary state
is different from that of the standard full S-brane.
Rather than corresponding to formation and decay of an unstable brane, it
corresponds to a brane decaying into closed strings
propagating in opposite directions in the covering space, as depicted
in Figure 2 ($X^0,\tilde X^0 > 0$ coordinatize the two sheets).

%%%%%%%%%%%%%%%%%%%%%%%%%%%%%%%%%%%%%%%%%%
\begin{figure}
\begin{center}
\includegraphics[width=6cm, height=4cm]{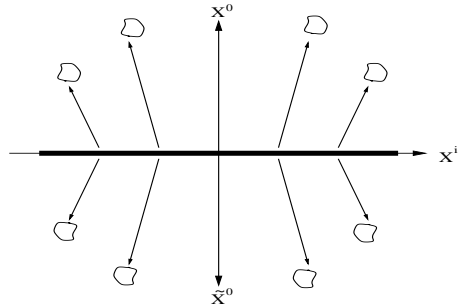}% Here is how to import EPS art
\end{center}
\caption{\label{fig:BBB_fig2}
Brane decay on the orbifold covering space.
}
\end{figure}
%%%%%%%%%%%%%%%%%%%%%%%%%%%%%%%%%%%%%%%%%%

On the fundamental domain (corresponding to the upper half space),
we then obtain a brane at the origin of time, decaying into closed
strings. This necessitates a prescription for preparing the brane
at the origin of time. Ref. \cite{Lambert:2003zr} proposed a Hartle-Hawking
contour integration prescription for spontaneously nucleating a
brane at $X^0=0$ which subsequently decays. We can use the same
prescription on the fundamental domain. On the covering space,
it corresponds to a double contour, with branches approaching the
origin along the imaginary time axes of $X^0,\tilde{X}^0$ and then
proceeding along real time axis into the opposite $X^0,\tilde{X}^0>0$
directions. The physical interpretation is otherwise the same as
in \cite{Lambert:2003zr}, except that now there is no spacetime before
$X^0=\tilde{X}^0=0$ -
the nucleation of the brane can truly be viewed as a "Big
Bang"-like event.
%The subsequent
The average total number density and total energy
density of the emitted closed strings can be calculated in a similar
manner as in \cite{Lambert:2003zr} for the Hartle-Hawking contour, with
similar results \cite{KKLN}.

If all the closed strings in the spacetime
originate from the brane decay, the latter would then serve
as an initial condition.
However, the spacetime is the covering space
of an orbifold, so it contains twisted sector strings as well.
How would they be connected with the decaying brane? In other
words, what is their role in the initial condition proposal -- can
the twisted strings also be associated with a decaying brane?

It turns out that on the spacetime orbifold there is another class
of S-branes or decaying branes to consider. We call these {\em
fractional S-branes}. They are constructed as follows.
Start again by the Wick rotation $X^0\mapsto iX$ to Euclidean
signature.
Formally in the bosonic theory one can also obtain an
array of D-branes at
$X=2\pi \sqrt{\alpha'}n$ by setting $\lambda =-\half$.
Tachyonic deformation
to $\lambda <0$ is not well defined in the bosonic theory because
it leads to the region where the tachyon effective potential is
unbounded from below. However, in superstring theory the effective
potential is bounded from below and symmetric about $T=0$, and
deformation to $\lambda =-\half$ is well defined. We will however
continue with the bosonic theory, in order to keep our proposal
clear -- we expect that our proposal can be extended to
superstrings. The boundary state of the decaying brane
is usually constructed by first compactifying
$X$ at selfdual radius, $X \sim X+2\pi \sqrt{\alpha'}$, then
the deformation acts as an $SU(2)$ rotation, and finally a projection
back to infinite radius is performed to obtain the final boundary
state.
Now consider the ${ Z}_2$ orbifold identification
where ${ Z}_2$ acts by
$X\mapsto -X$. Then $X=0$ becomes an orbifold fixed point. In the
array of D-branes at $X=2\pi \sqrt{\alpha'}n$, the brane at $X=0$
must be replaced by a fractional D-brane. If we compactify
on a selfdual radius as before, after the ${ Z}_2$ identification
to $S^1/{ Z}_2$ there are two fixed points, at $X=0$
and $X=\pi \sqrt{\alpha'}$.
Ref. \cite{Oshikawa:1996dj} studied the
conformal field theory of a free boson on this orbifold and found
that there are 8 fractional boundary states at the fixed point
$X_0=0,\pi \sqrt{\alpha'}$,
where the Dirichlet or Neumann untwisted sector boundary states
are combined with twisted sector boundary states.
Extending and elaborating a previous analysis by \cite{Recknagel:1998ih},
we have shown how the tachyonic deformation interpolates through the 8
fractional boundary states as $\lambda$ is varied. It is notable that
the process has a dual description in terms of a free boson on a circle --
without orbifold singularities -- where it corresponds to moving an
ordinary D-brane around the circle. We have also constructed a projection
back to the infinite radius.

We have also considered a more straightforward approach, starting
with the open string action with tachyonic deformations at the opposite
endpoints,
$$
  \delta S = -\lambda \int_{\partial_1 \Sigma} \cos (X/\sqrt{\alpha'})
           -\tilde{\lambda} \int_{\partial_2 \Sigma} \cos (X/\sqrt{\alpha'})
$$
and computed the annulus partition function with
the $Z_2$ reflection, using the
fermionization technique of \cite{Polchinski:1994my}. After reinterpreting
this in closed string variables, the result factorizes into
a form corresponding to
closed string propagation between two deformed boundary states.
The result is in agreement with the construction starting from the
selfdual radius. Details will appear in \cite{KKLN}.

%\section{Discussion}
{\em Discussion -- } We have considered S-branes on a spacetime
orbifold and shown that this leads to a new class of S-branes that
we have called fractional S-branes. We can see three interesting
directions for further study and potential applications. i) Our
construction is a toy model for a spacetime with a (fractional)
S-brane as the stringy initial state at the origin of time. In
particular, a space-filling brane decay provides homogeneous initial
conditions. Customarily homogeneity in brane cosmological models is
obtained by a collision of two branes which must be initially
aligned to be parallel, amounting to a careful fine-tuning. In our
proposal there is only a single brane. As the mathematical tools and
general understanding of the properties of S-branes and associated
spacetime solutions develops, we expect our construction to be one
step towards big-bang cosmological toy models where a decaying brane
at the finite origin of time leads to a holographic interpretation
of the spacetime that is created. ii) It would be interesting to
investigate our proposal in the context of string/brane gas
cosmology \cite{Brandenberger:1988aj}. Instead of a space-filling
infinite brane, one could start with all the spacelike dimensions
being compact. Initially the brane decay produces massive
(non-winding and winding) closed strings which
cascade to lighter modes, interact and presumably
thermalize in the end. The total energy of the system is the
initial mass of the unstable brane. The brane decay
might be used to "explain" the origin of the hot string gas which
drives the expansion of some of the compact directions.
iii) We can also speculate on
the relation of
our construction to the discussion of the emergence of the arrow
of time in \cite{Carroll:2004pn},
starting from a generic low-entropy state on an
"initial" spatial slice.
One of the issues with this
scenario was the nature of the initial state of the slice --
since the scenario leads to
arrows of time pointing away from the slice, we could argue that
the initial condition
could be chosen to be the fractional
S-brane that we constructed here.

\providecommand{\href}[2]{#2}\begingroup\raggedright\endgroup

\end{document}